\documentclass[a4paper]{jpconf}
\usepackage{graphicx}
\usepackage{amsmath,amssymb}

\begin{document}
\title{Carrier doping to a partially disordered state in the periodic Anderson model on a triangular lattice}

\author{Satoru Hayami$^1$, Masafumi Udagawa$^{1,2}$, and Yukitoshi Motome$^1$}

\address{$^1$Department of Applied Physics, University of Tokyo, Hongo 7-3-1, Bunkyo, Tokyo 113-8656}
\address{$^2$Max-Planck-Institut f\"{u}r Physik komplexer Systeme, N\"{o}thnitzer Stra\ss e 38, 01187 Dresden}

\ead{hayami@aion.t.u-tokyo.ac.jp}

\begin{abstract}
We investigate the effect of hole and electron doping to half-filling in the periodic Anderson model on a triangular lattice by the Hartree-Fock approximation at zero temperature. 
At half-filling, the system exhibits a partially disordered insulating state, in which a collinear antiferromagnetic order on an unfrustrated honeycomb subnetwork coexists with nonmagnetic state at the remaining sites. 
We find that the carrier doping destabilizes the partially disordered state, resulting in a phase separation to a doped metallic state with different magnetic order. 
The partially disordered state is restricted to the half-filled insulating case, while its metallic counterpart is obtained as a metastable state in a narrow electron doped region.
\end{abstract}

\section{Introduction}
Partial disorder is one of the peculiar orderings to geometrically-frustrated systems. 
It is stabilized by relieving the frustration with self-organizing the system into coexistence of magnetically-ordered sites and nonmagnetic sites.  
Interestingly, the partially disordered (PD) state has been observed in some metallic compounds~\cite{Ballou1998,Lacroix2010}, for example, in Laves compounds $R$Mn$_{2}$ ($R$=Dy, Tb, Th) and in $f$-electron compounds such as UNi$_{4}$B~\cite{Mentink, Oyamada} and CePdAl~\cite{Donni, Oyamada2}. 
In the Laves compounds, the PD state appears for Mn spins as a mixed state of ordered moments and nonmagnetic sites in the transient regime between magnetically-ordered and paramagnetic states while changing the lattice parameter.
On the other hand, in the $f$-electron  compounds, the PD state shows a character of a mixture of ordered moments and Kondo spin singlets. 
For this phenomenon, therefore, an instability toward  
Kondo singlet formation plays an important role. 
While the coexistence of ordered moments and spin singlets is expected to cause 
interesting magnetotransport phenomena, the comprehensive understanding of the PD state is not reached yet both experimentally and theoretically.  

In the present study, we focus on such a PD state with coexistence of magnetic ordering and nonmagnetic sites related with Kondo singlet formation. 
Such coexistence was theoretically studied as a consequence of competition between the Ruderman-Kittel-Kasuya-Yosida (RKKY) interaction~\cite{Ruderman,Kasuya,Yosida2} and the Kondo coupling~\cite{Kondo,Yosida1} under geometrical frustration. 
The PD state was indeed obtained by the mean-field (MF) approximation for a classical pseudomoment model~\cite{Regueiro, Lacroix} and by a sophisticated variational Monte Carlo method for the Kondo lattice systems~\cite{Motome}. 
Recently, the authors found a similar PD state in the periodic Anderson model on a triangular lattice at half-filling by the Hartree-Fock approximation~\cite{Hayami}. 
The PD state, however, is insulating, while it is metallic experimentally.

In this paper, to search for a metallic PD state, we extend the analysis of the PD state for the periodic Anderson model by hole and electron doping to half-filling. 
We will demonstrate that the PD state is immediately destabilized by carrier doping at the MF level, with showing a phase separation to a doped magnetically-ordered state. 
We show that a metallic PD state is obtained only as a metastable state in the MF solutions for electron doping.

\section{Model and method}
We consider a periodic Anderson model on a triangular lattice, whose Hamiltonian is given by

\begin{eqnarray}
{\mathcal{H}} = -t \sum_{\langle i,j \rangle, \sigma}(c^{\dagger}_{i\sigma} c_{j \sigma} + {\mathrm{H.c.}})
-V \sum_{i ,\sigma} (c^{\dagger}_{i \sigma} f_{i \sigma}+ {\mathrm{H.c.}})
 + U \sum_{i} f^{\dagger}_{i\uparrow}f_{i \uparrow} f^{\dagger}_{i \downarrow}f_{i \downarrow} + E_{0} \sum_{i, \sigma} f^{\dagger}_{i \sigma} f_{i \sigma},
 \label{Ham}
\end{eqnarray}
where $c^{\dagger}_{i \sigma}$ ($c_{i\sigma}$) and $f^{\dagger}_{i \sigma}$ ($f_{i\sigma}$) are the creation (annihilation) operators of conduction and localized electrons at site $i$ and spin $\sigma$, respectively.
The sum of $\langle i,j \rangle$ is taken over the nearest-neighbor sites on the triangular lattice. 
Hereafter, we take the hopping $t=1$ as an energy unit and the $c$-$f$ hybridization $V>0$. In the following calculations, we fix the Coulomb interaction between localized electrons $U=2$ and the atomic energy of localized electrons $E_{0}=-1$. 

In the previous study~\cite{Hayami}, the authors found a PD state at half-filling in the ground state of the model (\ref{Ham}) by applying the Hartree-Fock approximation to decouple the $U$ term with considering three-site unit cell. 
The PD state is stabilized between a 120$^\circ$ antiferromagnetic (AF) metal and a Kondo insulator (KI). 
Here, to explore a metallic PD state, 
we extend the analysis to the doped regions near half-filling by employing the same method as in Ref.~\cite{Hayami}. 
For better resolution, we take the sum over 400 $\times$ 400 grid points in the three-folded Brillouin zone in calculating the mean fields in this paper.

\section{Result and Discussion}

Figure 1 shows the result of the ground-state phase diagram including the carrier-doped regions near half-filling at $U=2$ and $E_0=-1$. 
Schematic figures for spin and charge states in each phase are also shown. 
$n$ is the electron density defined by $n=\sum_{i,\sigma} \langle c^{\dagger}_{i, \sigma}c_{i, \sigma} +  f^{\dagger}_{i, \sigma}f_{i, \sigma}\rangle/N $, where $N$ is the total number of sites. 
At half-filling $n=2$, a sequence of three phases --- an AF metal, PD insulator, and KI --- is obtained while increasing $V$, as in the previous study~\cite{Hayami}. 
In the PD state in the intermediate $V$ region, one of three sublattices becomes nonmagnetic, while a collinear AF order appears in the remaining two sublattices. 

When we dope carriers, the PD state is unstable to a phase separation (PS) for both electron and hole doping, as shown in figure~1. 
For electron doping, PS takes place between the PD state and the ferrimagnetic (F) metallic state. (In the very narrow window of $0.875 < n < 0.883$, PS takes place to the AF metal.) 
Meanwhile, for hole doping, PS occurs between the PD state and the AF metallic state extended from half-filling in the small $V$ region. 

The regions of PS are determined by the following procedure on the basis of grand canonical ensemble.  
Figure~\ref{PS}(a) shows the zero-temperature grand-canonical potential $\Omega =E-\mu n$ for various ordered states measured from that for the F state ($E$ is the internal energy per site and $\mu$ is the chemical potential). 
For each $\mu$, the lowest $\Omega$ state gives the ground state. 
Figure~\ref{PS}(b) shows the electron density $n$ as a function of $\mu$. 
In general, a transition between different ordered states is of first order, and hence, $n$ shows a discontinuity. 
The PS region is identified as the region where $n$ jumps. 
In the present example at $V=0.90$, the ground state changes from the AF state to the PD state at $\mu \sim 0.244$, and finally to the F state at $\mu \sim 0.391$, as shown in figure~\ref{PS}(a). 
Accordingly, $n$ changes discontinuously from $1.93$ to $2.00$ at $\mu \sim 0.244$ and from $2.00$ to $2.26$ at $\mu \sim 0.391$, as shown in figure~\ref{PS}(b). 
These two regions are PS: the former is between the AF metal and PD insulator, while the latter between the PD insulator and F metal.

\begin{figure}[t]
\includegraphics[width=67mm]{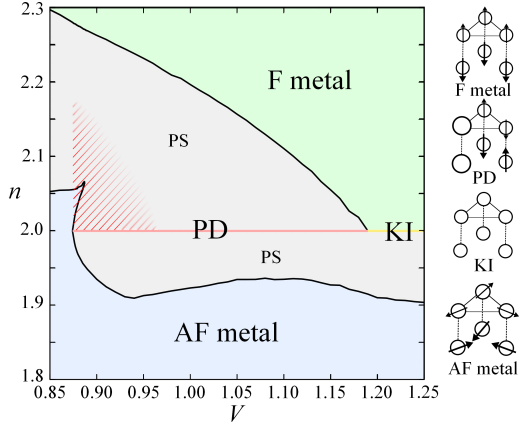} 
\hspace{5mm}
\begin{minipage}[b]{80mm}
\caption{Ground-state phase diagram of the model in eq.~(1) at $U=2$ and $E_{0}=-1$.
Schematic pictures for each phase are also shown:  
Upper (lower) circles show the $c$ ($f$) electron. 
The sizes of the circles represent local electron densities, and the arrows represent local spin moments, schematically. 
A hatched area shows the parameter region where a metallic PD state is obtained as a metastable state.  
}
\label{phase diagram}
\end{minipage}
\end{figure}

The PS regions are also identified by the canonical ensemble method. 
Figure~\ref{PS}(c) shows the internal energy $E$ as a function of $n$. 
In this approach, the state with the lowest $E$ gives the ground state for each $n$, and the PS regions are determined by drawing tangent lines for $E$ curves for different ordered states. 
At $V=0.90$, as shown in figure~\ref{PS}(c), two tangent lines can be drawn from the PD state at half-filling $n=2$: 
one is to the AF state at $n\sim1.93$ and the other is to the F state at $n\sim2.26$. 
These indicate two PS regions, consistent with the above results by the grand canonical ensemble scheme. 

We conducted similar analyses while varying $V$ to complete the phase diagram shown in figure~\ref{phase diagram}. 
As a result, we found that the PD state is destabilized by carrier doping to half-filling in the parameter range considered here.
That is to say, the PD state is limited to the insulating state at half-filling and a metallic PD state does not appear as a ground state within the present MF calculation. 

In contrast, the carrier doping to the AF state retains the AF order, and does not lead to PS. 
In particular, for hole doping, the AF metallic state extends widely, as shown in figure~\ref{phase diagram}. 
For electron doping, however, the AF state terminates at $n \sim 2.05$, and shows PS to the F state. 
On the other hand, the KI state changes into the F state without showing PS for electron doping, whereas it immediately exhibits PS to the AF state for hole doping. 

Although the PD state is unstable against the doping, we note that a doped metallic PD state is obtained as a metastable MF solution. 
The metastable metallic PD solution exists in the small region of $0.88 \lesssim V \lesssim 0.96$ for electron doping as indicated by the hatched area in figure~\ref{phase diagram}. The corresponding $\Omega$ and $E$ are shown in figure~\ref{PS}(a) and figure~\ref{PS}(c), respectively.
The proximity to a metallic PD state can be understood from the electronic structure at half-filling shown in figure~3 and figure~4 in Ref.~\cite{Hayami}. 
For $0.88 \lesssim V \lesssim 0.96$, the lowest unoccupied band (LUB) has an energy minimum around the M' point in the Brillouin zone, and has a considerable curvature in the band bottom. 
An electron doping to this band bottom retains the PD state in the doped region. 
On the other hand, in the larger $V$ region, the bottom of the LUB becomes flat with less curvature, and accordingly, the density of states (DOS) has a sharp peak at the upper edge of the gap. 
Doping to the flat band leads to an instability toward PS. 
Similar behaviors are seen in the hole doped case: 
For hole doping, DOS shows a sharp peak at the lower edge of the gap in the entire range of $V$, and hence, a metastable PD metal is not found there. 

Let us finally discuss a possibility to stabilize a metallic PD state on the basis of our results. 
In our results, the PD state is unstable against PS with the AF metal or the F metal. 
One may extend the PD state to doped regions by destabilizing the AF and F states with additional perturbation. 
A possibility is the AF Ising interaction between localized electrons, $I_z \sum_{\langle i,j \rangle} S_{i}^{z} S_{j}^{z}$, where $S_i^z = (f_{i \uparrow}^\dagger f_{i \uparrow} - f_{i \downarrow}^\dagger f_{i \downarrow})/2$. 
This term was shown to stabilize the PD state at half-filling in the previous study for the Kondo lattice systems~\cite{Motome}. 
Another possibility is the Coulomb interaction between conduction electrons. 
In particular, a long-range Coulomb interaction may prevent the system from PS. 
It will be interesting to examine whether a PD metal appears when PS is suppressed.
It is also important to consider other ordered states in larger unit cells beyond three-sublattice orderings in the present study. 

\begin{figure}[t]
\includegraphics[width=90mm]{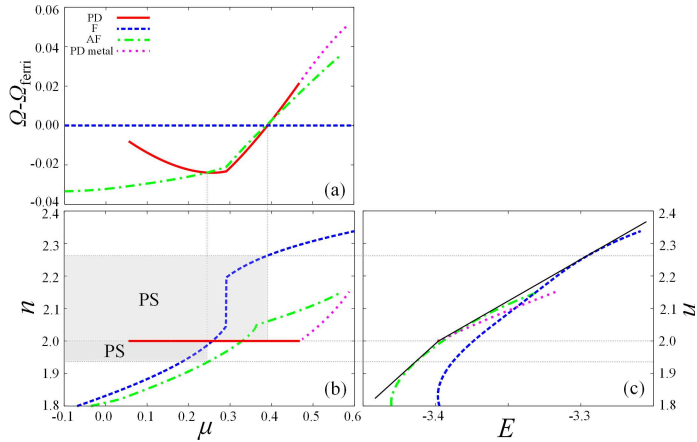} 
\hspace{3mm}
\begin{minipage}[b]{60mm}
\caption{
(a) Grand canonical potential at $T=0$ for each state measured from that for the F state as a function of the chemical potential $\mu$.
(b) Electron density $n$ as a function of $\mu$. The shaded regions show the PS regions.
(c) Internal energy per site $E$ plotted as a function of $n$. 
The straight black lines show tangent lines to determine the PS regions. 
All the data are at $U=2$, $E_0=-1$, and $V=0.90$. 
}
\label{PS}
\end{minipage}
\end{figure}

\section{Summary}
We have investigated the carrier doping to a partially disordered state in the periodic Anderson model on a triangular lattice by the Hartree-Fock approximation.
Elucidating the ground state phase diagram near half-filling, we found that hole and electron doping to the partially disordered phase leads to a phase separation to a doped magnetically-ordered state. 
In the small $c$-$f$ hybridization region, a metallic partially disordered state is obtained as a metastable solution for electron doping. 
We have discussed  possibilities to stabilize the metallic partially disordered state on the basis of the results.

We acknowledge helpful discussions with Takahiro Misawa, Junki Yoshitake, and Yutaka Akagi.
This work was supported by Grants-in-Aid for Scientific Research (No. 19052008, 21340090, 21740242, and 23102708), Global COE Program ``The Physical Sciences Frontier", from the Ministry of Education, Culture, Sports, Science and Technology, Japan.

\end{document}